# Nanoconfined circular and linear DNA – equilibrium conformations and unfolding kinetics

Mohammadreza Alizadehheidari[a], Erik Werner[b], Charleston Noble[c,d], Michaela Reiter-Schad[c], Lena K. Nyberg[a], Joachim Fritzsche[e], Bernhard Mehlig[b], Jonas O. Tegenfeldt[d], Tobias Ambjörnsson[c], Fredrik Persson[f] and Fredrik Westerlund[a]*

[a]Department of Chemical and Biological Engineering, Chalmers University of Technology, Gothenburg, Sweden
[b]Department of Physics, Gothenburg University, Gothenburg, Sweden
[c]Department of Astronomy and Theoretical Physics, Lund University, Lund, Sweden
[d]Department of Physics, Lund University, Lund, Sweden
[e]Department of Applied Physics, Chalmers University of Technology, Gothenburg, Sweden
[f]Department of Cell and Molecular Biology, Uppsala University, Uppsala, Sweden

**Abstract**

Studies of circular DNA confined to nanofluidic channels are relevant both from a fundamental polymer-physics perspective and due to the importance of circular DNA molecules *in vivo*. We here observe the unfolding of DNA from the circular to linear configuration as a light-induced double strand break occurs, characterize the dynamics, and compare the equilibrium conformational statistics of linear and circular configurations. This is important because it allows us to determine to which extent existing statistical theories describe the extension of confined circular DNA. We find that the ratio of the extensions of confined linear and circular DNA configurations increases as the buffer concentration decreases. The experimental results fall between theoretical predictions for the extended de Gennes regime at weaker confinement and the Odijk regime at stronger confinement. We show that it is possible to directly distinguish between circular and linear DNA molecules by measuring the emission intensity from the DNA. Finally, we determine the rate of unfolding and show that this rate is larger for more confined DNA, possibly reflecting the corresponding larger difference in entropy between the circular and linear configurations.

**Introduction**

Nanofluidic channels have, during the last decade, emerged as a powerful tool to study single DNA molecules using fluorescence microscopy.[1,2] When a DNA molecule is confined to a channel that is narrower than its radius of gyration it must stretch along the channel. The resulting extension of the DNA molecule depends linearly upon the contour length.[3] Furthermore, there is no fundamental upper limit of the size of the DNA that can be investigated. A main advantage of using nanochannels, compared to most other single DNA-molecule techniques, such as optical[4] and magnetic[5] tweezers, is the possibility to stretch DNA without attaching any handles. This allows us to straightforwardly stretch circular DNA and DNA configurations with higher topologies.

The polymer physics of confined DNA molecules has been studied intensively during the last decade[6,7] and the effects of solvent characteristics[8,9] and molecular crowding[10,11] have been analyzed in detail. Recently, several groups have used nanofluidic channels to investigate the physical properties of nanoconfined DNA-protein complexes[12-16] and for optical mapping of single DNA molecules[17-21].

Circular DNA is of interest for at least two reasons. First, several different biological DNA molecules are circular, such as mitochondrial DNA in eukaryotic cells and chromosomal and plasmid DNA in bacteria. The latter has recently attracted strong interest since plasmids carry genes involved in antibiotic resistance.[22] Second, it is of interest to analyze the equilibrium conformational fluctuations of confined circular DNA because it makes it possible to test statistical theories for the extension of confined polymers in new ways. Segments of strongly confined circular DNA configuration may interact substantially, even when they are located far from each other along the contour length.[7]

A peculiar property of DNA stained with fluorescent dyes, such as YOYO-1 (the dye used here), is that the dye in its excited state forms reactive oxygen species that cause single-strand breaks on DNA, so called "nicks".[23,24] When two such single-strand nicks occur sufficiently close to each other, but on opposite strands, then the duplex DNA between the two nicks becomes unstable and a double-strand break occurs. Since the folded conformation of the now linear configuration is entropically unfavorable, the broken DNA must unfold. In our experiments we observe how this unfolding proceeds in real time. After sufficiently long time a new equilibrium is reached, making it possible to compare the equilibrium conformational fluctuations of the circular and linear configurations of the same DNA molecule. It is much easier to compare to theoretical predictions for equilibrium conformational fluctuations of circular and linear DNA configurations when their contour lengths are the same.

The channel sizes that we use in our experiments are of the same order as the persistence length of DNA. As a consequence, the conformational statistics of the confined DNA molecule cannot be adequately described by any of the established theories for confined polymers. However, we here show that the ratio between the extension of the circular and linear topologies of a 42 kbp DNA falls between theoretical predictions for the Odijk regime, valid at very strong confinement, and the extended de Gennes regime, valid at weaker confinement. As would be expected from a simple interpolation, the more confinement increases and buffer concentration decreases, the closer are the experimental results to the prediction for the Odijk regime. We also demonstrate that the rate of unfolding is higher for circular DNA at lower buffer concentrations where it is more extended, in agreement with the theoretical prediction that the difference in free energy between the two states is larger. In our experiments we find that circular DNA molecules are more likely to break very close to the ends of the extended circle and potential reasons for this bias are discussed. Finally, we show that measuring the local emission intensity of the confined DNA molecule allows us to automatically separate circular and linear DNA molecules in a nanofluidic device.

**Materials and Methods**

DNA samples were mixed in TBE buffer (Tris-Borate EDTA) prepared by dissolving and diluting a standard 10x TBE tablet (Medicago) in milli-Q water to the desired concentration. The reducing agent β-mercaptoethanol (Sigma-Aldrich) was added to the sample (3% v/v) to reduce photo-induced breaking of the DNA. The experiments were performed with circular charomid DNA (9-42, 42.2 kbp) purchased from Wako (Nippon Gene). The dimeric cyanine dye YOYO-1 from Invitrogen was used to stain DNA in dye:bp ratios of 1:10 , 1:20 and 1:40. The samples were not equilibrated, in order to allow different binding ratios to be investigated.[25]

The nanofluidic chips were fabricated in fused silica as described in Ref. [1]. Electron-beam lithography and reactive-ion etching were used to define the nanochannels with the following approximate dimensions: channel width 100 nm, channel depth 100 nm or 150 nm, channel length 500 μm. Microchannels with a width of approximately 50 μm and a depth of circa 1 μm connect the nanochannels to four reservoirs of the device. The DNA sample was loaded into the one of the reservoirs and brought into the nanochannels by a pressure-driven flow.

For DNA imaging, an epi-fluorescence microscope (Zeiss AxioObserver.Z1) equipped with a high quantum yield EMCCD camera (Photometrics Evolve) and a 100x oil immersion objective with high numerical aperture (NA 1.46) from Zeiss was used. A stack of 200 images was recorded for each DNA molecule with an exposure time of 100 ms per frame at a rate of 7 frames per second.

Data analysis was performed with the program ImageJ (http://rsbweb.nih.gov/ij/) and algorithms

written in Matlab®. In short, DNA molecules were marked in the image stack by a region of interest and the algorithm computed an intensity trace of each individual molecule by averaging over the region of interest in each frame. A kymograph (time trace) was then constructed by stacking these intensity traces on top of each other. Examples are shown in Fig. 1. Each line in the kymograph corresponds to one frame in the image stack. For each frame the intensity trace was fitted to an error-function profile. This procedure gives the average fluorescence intensity and the extension of the DNA molecule in each frame. Note that the above refers to the profile along the channel direction, averaged over the transverse direction. The form of the transverse density profile was discussed by Werner *et al.* in [26].

From the kymograph (Fig. 1A) it is possible to determine the time when the circular DNA broke, and to find the position of the corresponding breaking point in the channel. To determine the break time and position we first smoothed the kymograph aligned by the intensity weighted center (Fig. 1B) using two-dimensional averaging. Using Otsu's method[27] we distinguished three different regions in the kymograph (Fig. 1B), corresponding to circular and linear DNA configurations, and a background region. These three regions differ in their fluorescence-emission intensities and are identified using a Matlab® script. In Fig. 1B the light regions with the highest fluorescent intensities pertain to times and positions in the kymograph corresponding to unbroken circular DNA configurations. The somewhat darker regions correspond to broken linear DNA configurations. The background is black. In a second step we located the triangular kymograph area that corresponds to broken DNA in the process of unfolding (visible in panel B of Fig. 1). The coordinates of the highest corner of this triangle give the breaking position and the breaking time. All results were verified by manual inspection. More details on the procedure can be found in the Supporting Information.

A Matlab® script was used to automate the analysis of the unfolding process. Fig. 2 shows the trace of molecule extension as a function of time, overlaid with a fitted empirical formula given by

$$R(t) = 1/2\ (R_{lin} + R_{circ}) + 1/2\ (R_{lin} - R_{circ})\mathrm{erf}((t-t_0)/\tau) \qquad (1)$$

From the fit the start and end points of the unfolding process were determined as the first points where the difference between adjacent time points in the fitted function differs by more than a user-set threshold, here chosen to be 0.2 pixels (32 nm). The rate of unfolding is taken to be the slope of the line connecting the start and end points (black line in Fig. 2). In addition, the equilibrium extension and its standard deviation for the circular and linear configuration of the DNA were calculated by averaging over all values of R(t) before the start point and after the end point, respectively.

## Results

We start by describing equilibrium extensions of circular and linear DNA configurations. First we show that it is possible to experimentally detect when and how circular DNA breaks and unfolds to the linear configuration, using the fact that there are apparent differences in extension and emission intensity of the circular and the linear configurations (Materials and Methods). Figure 1A shows two examples of extracted kymographs for initially circular DNA configurations that break at the center (left) and the end (right), respectively, and subsequently unfold. Figure 1B shows the same data but aligned as described in the Methods section. Figure 1C compares snap-shots of a YOYO-1-stained circular DNA before and after unfolding in the nanochannel. As expected, the DNA is more extended when it has reached its linear equilibrium configuration. The data in 0.05X TBE buffer is collected in 100x100 nm$^2$ channels and all other data is collected in 100x150 nm$^2$ channels. The sample in 0.05X TBE is included in order to push the extension as close to the Odijk regime as possible (see below). Furthermore, when labeling the DNA with YOYO-1 we *do not* equilibrate our samples. This makes it possible to investigate a wide range of dye loads.[25]

We now compare the experimental observations with theoretical predictions. A DNA molecule is commonly modeled as a worm-like chain with contour length $L$, persistence length $l_p \approx 50$ nm, and effective width $w_{eff} \approx 5$–20 nm (depending on buffer concentration).[1] Since the channel dimensions $D \approx 100$–150 nm are of the same order as the persistence length of the DNA, there are no exact theories for the extension ratio. However, we show below that the experimental results are consistent with a simple interpolation between the expected results for the Odijk regime and the extended de Gennes regime (valid for smaller and larger channels, respectively).

### *Theory for the extended de Gennes regime*

For a semi-flexible polymer, such as DNA, there exists a parameter regime at intermediate channel sizes $l_p < D < l_p^2/w_{eff}$, known as the extended de Gennes regime.[28-30] In this regime, an asymptotically exact theory for the equilibrium statistics of a linear polymer predicts that the mean and variance of the extension are given by [31]

$$R_{lin} = 1.18L \left(\frac{l_p w_{eff}}{D^2}\right)^{\frac{1}{3}} \quad (2)$$

$$\sigma_{lin} = 0.51(Ll_p)^{1/2} \quad (3)$$

Unfortunately it is not obvious how to generalize this model to circular polymers. We expect that the method described in Ref. [30] may be applied to circular DNA, but this has not yet been shown. An alternative way to derive the scalings (though not the prefactors) of Eqs. 2-3 is to estimate the R-dependence of the free energy by a Flory-type mean-field argument, yielding[31, 32]

$$\frac{F_{lin}(R_{lin})}{kT} = A\frac{R_{lin}^2}{Ll_p} + B\frac{L^2 w_{eff}}{R_{lin}D^2} \quad (4)$$

$$R_{lin} = L\left(\frac{Bl_p w_{eff}}{2AD^2}\right)^{\frac{1}{3}} \quad (5)$$

$$\sigma_{lin}^2 = \frac{Ll_p}{12A} \quad (6)$$

Here, A and B are prefactors of order unity. This calculation for the free energy is easily generalized to a circular polymer, by treating it as two linear chains of length $L/2$ that are forced to overlap:

$$\frac{F_{circ}(R_{circ})}{kT} = 2A\frac{R_{circ}^2}{(L/2)l_p} + B\frac{L^2 w_{eff}}{R_{circ}D^2} \quad (7)$$

$$R_{circ} = L\left(\frac{Bl_p w_{eff}}{8AD^2}\right)^{1/3} \quad (8)$$

$$\sigma_{circ}^2 = \frac{Ll_p}{48A} = \sigma_{lin}^2/4 \quad (9)$$

Here $A$ and $B$ are the prefactors mentioned above. Comparing Eqs. 5-6 and Eqs. 8-9, one finds

$$R_{lin}/R_{circ} = 1.59 \quad (10)$$

$$\frac{\sigma_{lin}}{\sigma_{circ}} = 2 \quad (11)$$

It should be noted that whereas the circular DNA molecule is presumably in an unknotted state, the theory averages over all possible knotting states. To which extent this matters remains to be tested.

***Theory for the Odijk regime***

The Odijk regime[33] applies to polymers confined to channels with a diameter smaller than the persistence length. In this regime, the mean and variance of the extension of a linear DNA molecule

confined to a rectangular channel with cross section $D_x \times D_y$ are given by

$$R_{lin} = L\left[1 - \alpha \frac{D_x^{2/3} + D_y^{2/3}}{l_p^{2/3}}\right] \quad (12)$$

$$\sigma_{lin}^2 = \beta L \frac{D_x^2 + D_y^2}{l_p} \quad (13)$$

where $\alpha = 0.09137 \pm 0.00007$, $\beta = 0.00478 \pm 0.00010$.[34] As in the extended de Gennes regime, circular DNA in the Odijk regime can be treated as two strands going in opposite directions. Calculating the statistics of the extension is complicated by the interaction between the two strands. Yet the extension can be bounded above and below by considering the two extreme cases of either no interaction, or complete separation such that each strand only explores half of the channel. For a square channel this analysis yields that the mean and variance of the extension are bounded by

$$L/2\left[1 - 2\alpha \frac{D^{2/3}}{l_p^{2/3}}\right] \leq R_{circ} \leq L/2\left[1 - 1.63\alpha \frac{D^{2/3}}{l_p^{2/3}}\right] \quad (14)$$

$$\beta L \frac{5D^2}{16 l_p} \leq \sigma_{circ}^2 \leq \beta L \frac{D^2}{2 l_p} \quad (15)$$

Here D is the side length of the square. Comparing Eqs. (12) and (14) yields a ratio of extensions in the interval $R_{lin}/R_{circ} = 1.92 - 2$ at $D = l_p$, approaching 2 as $D$ tends to zero. Comparing instead Eqs. (13) and (15) shows that the ratio of standard deviations must lie in the interval $\sigma_{lin}/\sigma_{circ} = 2 - 2.5$, throughout the Odijk regime

To summarize, we expect an extension ratio $R_{lin}/R_{circ}$ of 1.59 in the extended de Gennes regime and 1.92 - 2 in the Odijk regime, and a fluctuation ratio $\sigma_{lin}/\sigma_{circ}$ of 2 in the extended de Gennes regime and 2 - 2.5 in the Odijk regime.

***Experiments***

Figure 3 and Table 1 show experimentally measured equilibrium extensions of circular and linear configurations for different buffer concentrations. For both linear and circular configuration we see that the extension increases with increasing confinement and decreasing buffer concentration. We also observe that the relative extension of the circular configuration is higher than the linear configuration,

*i.e.* the extension ratio is always smaller than 2.

All experimental data consistently falls between the two extremes predicted by the theories (Eqs. 6 and 9). Furthermore, as the degree of confinement increases and the buffer concentration decreases, the extension ratio gradually shifts from the prediction for the extended de Gennes theory towards that for the Odijk theory, as expected. The spread within each data set is largest at the lowest buffer concentration. This is expected too since the YOYO-1 dye used for staining the DNA affects the extension most at low buffer concentrations, and the samples used were not equilibrated.[25]

Figure 4 and Table 1 show equilibrium extension fluctuations for the circular and linear configuration of the same DNA molecule. Although the spread is much larger than for the extension ratio in Figure 3, there is a clear trend that the fluctuations decrease with decreasing buffer concentration. This is consistent with the theoretical prediction that fluctuations are smaller in the Odijk regime. We consistently observe an experimental ratio above 2, which also increases as we move from the extended de Gennes regime (ratio = 2) towards the Odijk regime (ratio ≥ 2).

A comment on how we extract the DNA extension and its variance from experiments are in order. The common approach is to fit the individual intensity traces in the kymograph to error-function windows.[1] There is no microscopic justification for using error functions, but this procedure works very well and we have adopted it here (see Materials and Methods). It provides time-resolved estimates of the extension. Each intensity trace is averaged over a small time window, we have neither accounted for the corresponding motion blur, nor for diffraction corrections. An alternative way of extracting an average extension is to time average the center-aligned kymograph (Fig. 1B) and to fit the resulting intensity profile. A detailed theoretical calculation shows that the extensional fluctuations give rise to an error-function with a window size which can be explicitly expressed in terms of $\sigma_{lin}$ and $\sigma_{circ}$, see equations (S.11) and (S.12) in the SI. It is important to note that the functional form of the profile of the time-averaged intensity results from extensional fluctuations, unrelated to the processes that determine the functional form of the time-resolved intensity profile.

From a biotechnological perspective it is of interest to distinguish, in an automated fashion, whether a DNA fragment is made up of DNA with linear or with circular configuration. Figure 5 shows the ratio of the mean fluorescence intensity and length between circular and linear configurations at different buffer concentrations. We define the intensity per length ratio as the integrated fluorescence intensity of the kymograph of each molecule, divided by the extension of the DNA for the circular and linear states, respectively. The ratio is close to 2 at all buffer concentrations, as expected, since the double-folded circular DNA contains 1.6-1.9 times more DNA per pixel, as demonstrated in Table 1. That the extension ratio decreases with increasing buffer concentration, as discussed above, explains why the emission ratio is lowest at the highest buffer concentration. This concludes our discussion of the equilibrium extensions of circular and linear configurations of the same DNA molecule confined to a channel.

We now discuss the unfolding dynamics. As described in the introduction, a circular DNA molecule must unfold after the DNA circle is broken. From the kymographs we can extract the rate of unfolding as described in Materials and Methods. Figure 6 compares the mean values of the rate of unfolding at different buffer concentrations. We observe that the rate of unfolding decreases as the buffer concentration increases (and the extension of the DNA decreases). Next, we determine whether the rate of unfolding depends on the position of the double strand break along the contour. We divide the molecules into two categories, one for all the molecules where the break occurs in the central 50% of the molecule ("center") and one for the molecules where the break occurs in the outer 50% ("end", see Figure 6 (top) for a schematic illustration). Figure 6 (bottom) shows that the rate of unfolding is higher if the DNA breaks in the center of the molecule than if it breaks in the end. The unfolding process is fast, therefore it is difficult to determine unfolding rates, leading to large error bars for our rate estimates.

Interestingly, we note that unfolding originating at one of the very ends is significantly over-represented at all buffer concentrations (Figure 7). Between 30% and 50% of the molecules start to unfold at the very end of the molecule (outmost 2%-10%, depending on buffer concentration). Moreover, the tendency for circular DNA to unravel from the end increases with increasing buffer concentration. The fact that center regions are equally represented at all buffer concentriond is a good confirmation that our program for detecting nicking positions works and that the effect we see is a true effect.

**Discussion**

By measuring the average extension before and after unfolding of individual circular DNA molecules undergoing photo-damage, we investigate how the ratio of the equilibrium extension of the two configurations varies with confinement and buffer concentration. Since we measure the extension of the *same* molecule in the circular and linear configuration we avoid any bias due to that the linear DNA has been fragmented. A circular DNA molecule is by default intact and a linear DNA molecule originating from a circular DNA molecule must, at least initially, contain the same number of base-pairs. The ratio between the linear and circular extensions increases from $1.64 \pm 0.07$ at 2.5X TBE

(weaker confinement) up to 1.84 ± 0.07 at 0.05X TBE (stronger confinement) (see Table 1). Since a circular DNA molecule is akin to two linear molecules connected at both ends, an extension ratio below two suggests that the extension of the circular molecule is increased by repulsive interactions between the two overlapping linear strands that make up the circular DNA. In other words, as the channels become smaller compared to the persistence length of DNA, excluded volume effects become less important. The measured ratios lie between the values predicted for the Odijk and the extended de Gennes regimes. Values obtained by measurements at high buffer concentration (weaker confinement) are closer to the value predicted for the extended de Gennes regime, and *vice versa*.

Experimental comparisons between circular and linear DNA configurations have been performed before. Levy *et al.* investigated partly folded linear DNA inserted into nanochannels.[35] They performed their experiments at high buffer concentration (5X TBE) and in wider channels than in our experiments. Their ratio between the extensions of the unfolded and folded regions is lower than in our experiments (Table 1), in agreement with the fact that their experiments were done at conditions closer to the extended de Gennes regime. Furthermore, Levy *et al.* observed that the ratio between the measured equilibrium extensions for circular and linear DNA decreases when the degree of confinement decreases, again in agreement with our data. Similar behavior was observed by Persson *et al.* who compared linear and circular DNA at different degrees of confinement in nanofunnels.[7] They showed that the extension increases more rapidly with increasing confinement for linear than for circular DNA. This implies that the ratio between the extensions increases as the degree of confinement increases. This conclusion agrees with our experimental and theoretical results, since decreasing the channel size and decreasing the buffer concentration both decrease the ratio $D / l_p$.

In our experiments the ratio between the standard deviations of the extensions of the circular and linear configurations increases from 2.26 ± 0.52 at 2.5X TBE to 2.64 ± 0.64 at 0.05X TBE (and stronger confinement). The latter value is slightly larger than our prediction for the Odijk regime allows (2.5), yet this small discrepancy could potentially be explained by the fact that for given external conditions, the extension of the circular configuration is closer to its full extension than the linear configuration. The circular DNA thus approaches the low-fluctuation Odijk regime at a weaker confinement than the linear DNA. This interpretation is supported by results of simulations,[36] showing that the Odijk regime extends to larger channel sizes for circular polymers, as compared with linear polymers.

In Figure 5 we demonstrate that the emission intensity from YOYO-labeled DNA is approximately two times higher for the circular configuration than for the linear one. The ratio between the intensities is lowest at the highest buffer concentration, in excellent agreement with the observation that the extension ratio decreases with increasing buffer concentration. Our results, together with the fact that we have established conditions for uniform DNA staining,[25] imply that it is possible to automatically differentiate linear and circular DNA in nanofluidic chips by calibrating the emission intensity and using an intensity cut-off. We believe this to be a promising complement to for example existing gel-electrophoresis techniques. The advantages with this new nanofluidic approach include speed, low sample consumption, and the possibility to identify molecules that appear at very low numbers. The nanochannel measurements are also directly compatible with Lab-on-a-chip like devices.

The data collected also allows us to analyze the rate of unfolding. Figure 6 shows that the rate of unfolding increases with increasing extension of the DNA. This could be due to the fact that the difference in free energy between the circular and linear configuration increases with increasing extension. In other words, there is a larger driving force for unfolding, which leads to a faster unfolding rate. Furthermore, we find that the average unfolding rate depends on the position of the breaking point.. We observe a trend that, at all conditions studied, the unfolding rate is higher when the break occurs in the center of the blob. This is what one would expect assuming that it takes more time to unfold a long segment than a short one. Unfortunately, the relatively short contour length of the DNA used in our experiments leads to a large experimental uncertainty in the rate, making it impossible to draw any quantitative conclusions from the data.

In Figure 7 we observe that a surprisingly large fraction of molecules (30-50%) appear to break at the very end of the blob. There are several possible explanations of this phenomenon. It is important to remember that double strand breaks occur when two single strand nicks, on opposite strands, occur so close that the duplex between them becomes unstable. One possible explanation for the observed bias is that the minimal distance between nicks is larger at the ends of the molecule, where the polymer must bend to turn back. However, we can rule out this effect by observing that the fraction of end-unfolding molecules increases with increasing buffer concentration. If this explanation were true, the bias should be larger at low buffer concentration where the DNA is less flexible, but this is not observed. Furthermore, the channels are so wide that the energy required to make this bend is small. A second possible explanation is that the bias is due to a higher local DNA concentration in the ends of the blob, but this is not observed in any of the data collected (data not shown). A third possible explanation is that friction forces cause the "head-end" to split as the molecule diffuses in the channel. This would lead to most of the end-unfolding events observed at the "head-end" of the DNA molecule, but this was not observed (data not shown). A fourth possibility is that oxygen concentration is lower in the region of the channel occupied by the DNA than in the rest of the channel, which could lead to a higher concentration of reactive oxygen species at the end of the DNA, and thus higher nicking rate near the end of the blob. Further experiments are required to determine the cause of this curious phenomenon.

To conclude, we have compared the polymer physics of two different DNA configurations, circular and linear, of the same DNA molecule. This is possible since a DNA with a circular configuration spontaneously unfolds to the linear form when a double strand break occurs. We show that the statistics of the equilibrium extensions gradually change from agreeing with our prediction for the extended de Gennes regime towards agreeing with the expectation for the Odijk regime, as the relative confinement increases. The rate of unfolding also increases as the stretching increases, consistent with a larger difference in free energy at high stretching. Finally, we show that the double strand break required for the unfolding occurs at the very end of the extended circle to a much larger extent than expected.

**Acknowledgements**

Dr. Jason Beech is acknowledged for fruitful discussions. This project was funded by a grant to F.W. from Chalmers Area of Advance in Nanoscience and Nanotechnology. J.O.T. is supported by the Swedish Research Council (VR) grant no. 2007-584. T.A. is supported by the Swedish Research Council, grant no. 2009-2924 and The Carl Tryggers Foundation, grant no. 12:13. B.M. is supported by the Swedish Research Council and the Göran Gustafsson Foundation for Research in Medicine and the Natural Sciences.

**Figures**

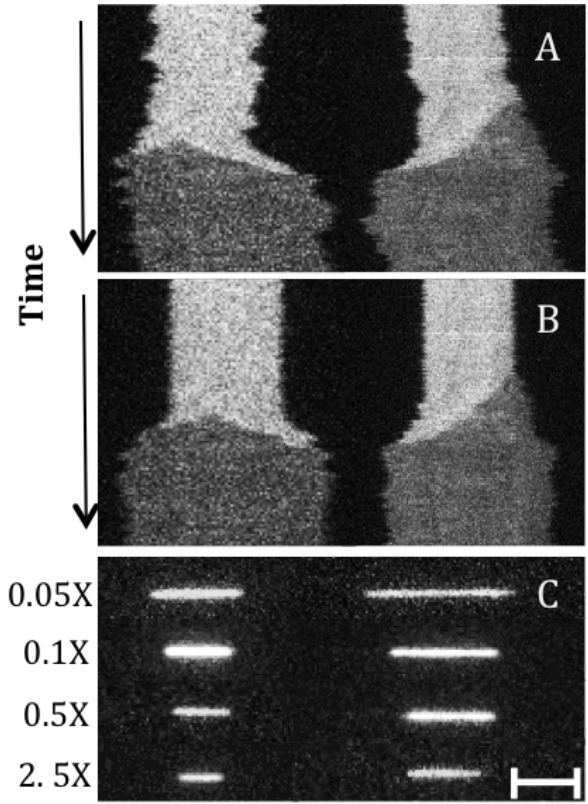

***Figure 1.*** *Kymographs, i.e. time (y-axis) vs. extension (x-axis), showing unfolding of circular DNA: (A) Raw-kymographs, (B) kymographs aligned by the center of mass. Left: Circular DNA unfolding from the center. Right: Circular DNA unfolding from the end. (C) Snap-shots of YOYO-1-stained circular DNA before (left) and after (right) unfolding at four different buffer concentrations in channels with dimensions of about $150\times100$ nm$^2$ for 0.1X TBE, 0.5X TBE and 2.5X TBE solutions, and $100\times100$ nm$^2$ for 0.05X TBE. The scale-bar corresponds to 5μm in all images.*

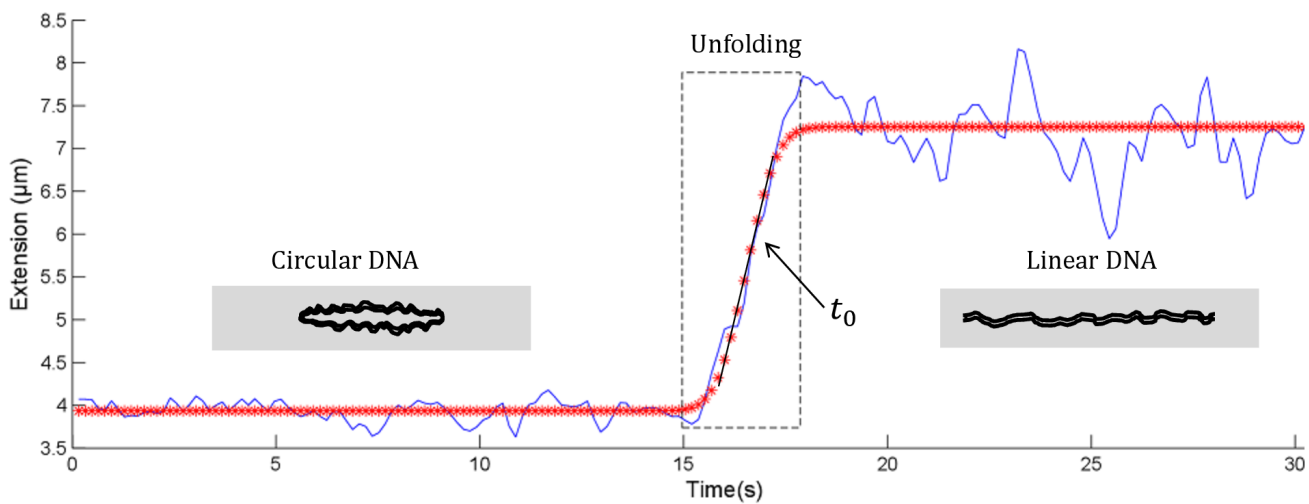

*Figure 2. Analysis of the unfolding of a DNA molecule as a function of time. The blue line corresponds to the experiment while the red dots correspond to the best fit of Eq 1 to the data. The slope of the black line corresponds to the average rate of unfolding. Insets: Illustrations of the circular and linear equilibrium configurations of the DNA molecule,*

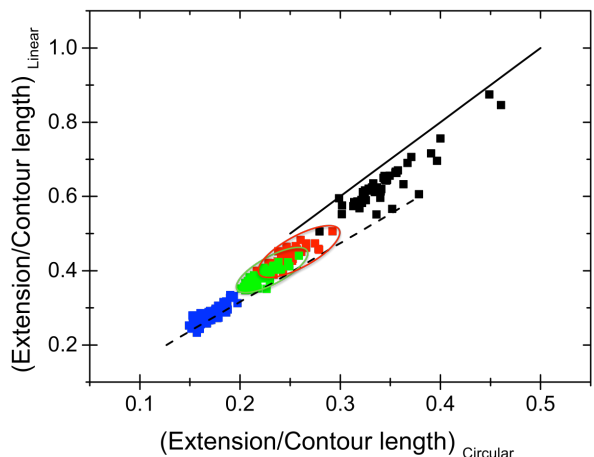

**Figure 3.** *Equilibrium extension divided by contour length for circular (x-axis) and linear (y-axis) configurations of charomid DNA at four different buffer concentrations. Each data point represents a single molecule confined to 100x150 $nm^2$ channels in 2.5X TBE (blue squares), 0.5X TBE (green squares), and 0.1X TBE (red squares). Black squares correspond to 100x100 $nm^2$ channels in, 0.05X TBE. The red and green circles emphasize the slight overlap between the data points at 0.5X and 0.1X TBE. The contour length is calculated assuming 1 YOYO-1 per 10 bp. The dashed line corresponds to the expected value in the extended de Gennes regime ($R_{lin}=1.59R_{circ}$, see Eq. 10). The solid line corresponds to the Odijk regime, approached as the confinement tends to zero ($R_{lin}=2R_{circ}$, see Eq. 14).*

*Note that the lines extend outside the ranges of validity of the theories described above. For each buffer concentration, between 50 and 85 molecules were imaged.*

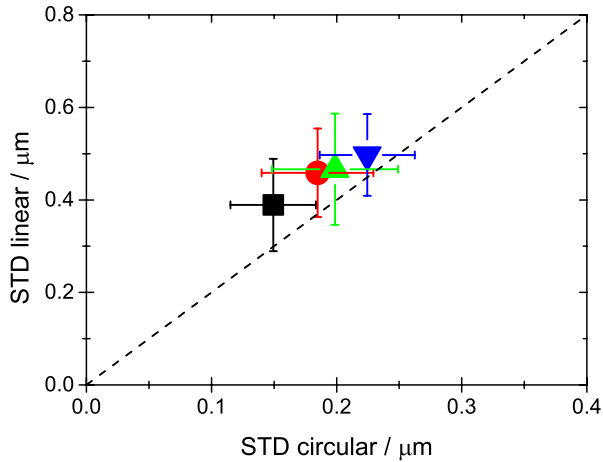

***Figure 4.*** *Standard deviations of the equilibrium extensions of circular and linear configurations of charomid DNA at four different buffer conditions. Symbols represent the average of all data points at a certain condition with error bars corresponding to the standard deviation: 2.5X TBE (blue), 0.5X TBE (green), and 0.1X TBE (red) in 100x150 nm² channels, and 0.05X TBE (black) in 100x100 nm² channels. The dashed line corresponds to $\sigma_{lin}=2\sigma_{circ}$ (see last paragraph in the Subsection Theory for the Odijk regime and Table 1). For each buffer concentration, 50-85 molecules were imaged.*

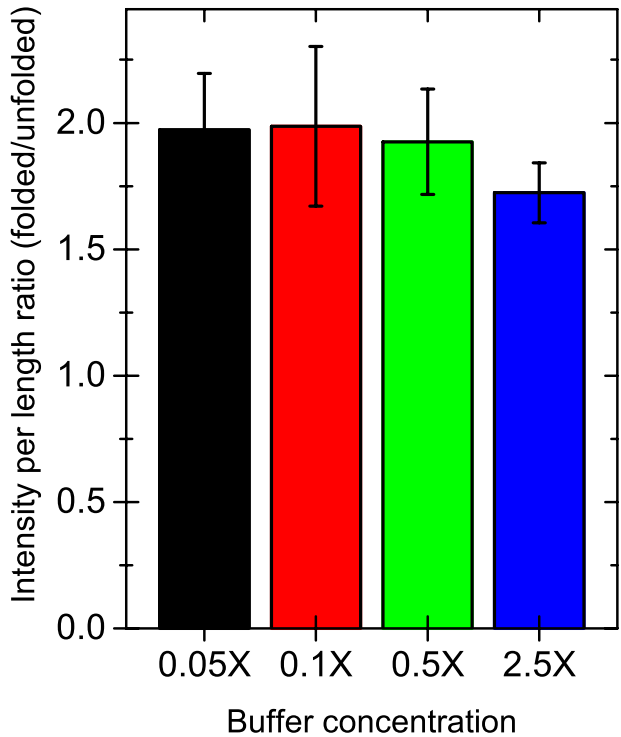

***Figure 5.*** *Histogram of the mean intensity-per-length ratio of circular and linear DNA configurations at four different buffer concentrations: 0.05X TBE (black) in 100x100 nm² channels and 0.1X TBE*

(red), 0.5X TBE (green) and 2.5X TBE (blue) in 100x150 nm$^2$ channels. Error bars show standard deviations. For each buffer concentration, 50-85 molecules were imaged.

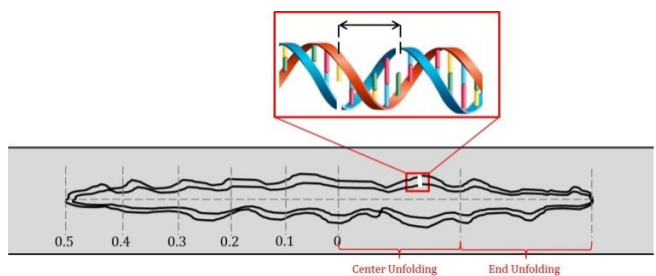

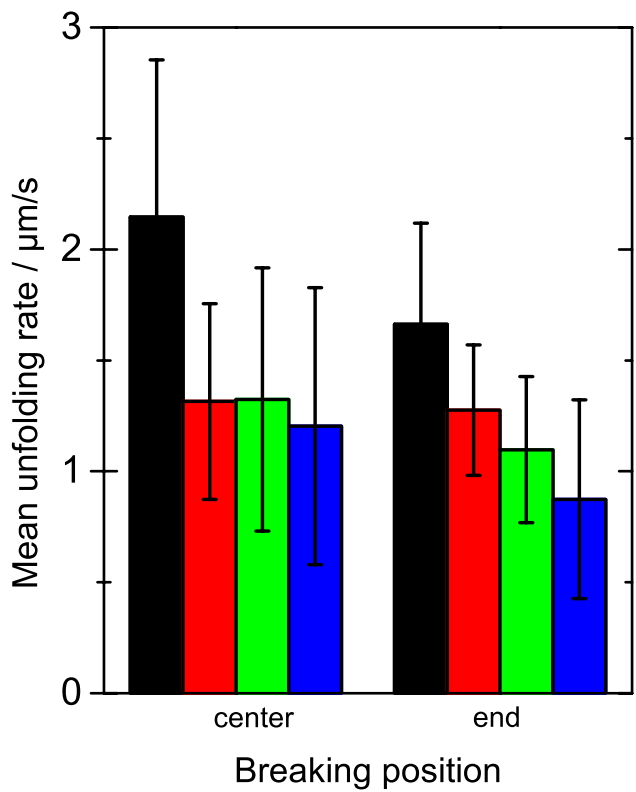

*Figure 6.* Top: cartoon of a nanoconfined circular DNA molecule, showing different regions. "Center" corresponds to unfolding starting in the mid 50% of the circular DNA blob while "End" corresponds to the outer 50% of the circular DNA blob. The close-up illustrates the possibility of two single-strand nicks occurring close enough on opposite strands to cause a double-strand break. Bottom: Histogram of the mean unfolding rate of circular DNA at different buffer concentrations: 0.05X TBE (black) in 100x100 nm$^2$ channels and 0.1X TBE (red), 0.5X TBE (green) and 2.5X TBE (blue) in 100x150 nm$^2$ channels. Error bars show standard deviations. For each buffer concentration, 50-85 molecules were imaged.

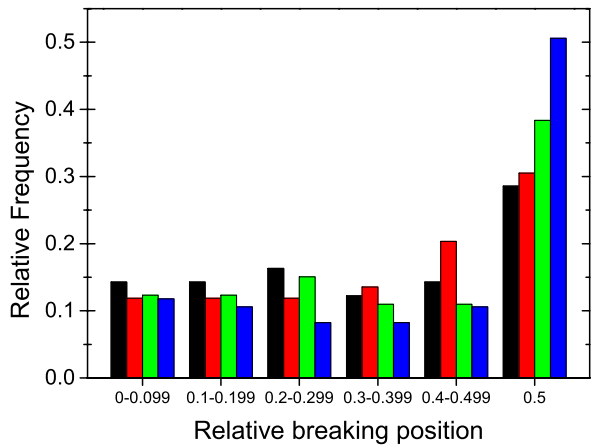

*Figure 7.* *Histogram of the breaking position of circular DNA at four different buffer concentrations: 0.05X TBE (black) in 100x100 $nm^2$ channels and 0.1X TBE (red), and 0.5X TBE (green) in 100x150 $nm^2$ channels. For each buffer concentration, 50-85 molecules were imaged.*

# Tables

**Table 1.** *Average extension ratios and standard-deviation ratios at four different buffer concentrations. The data from Levy et al [35] is given in terms of a parameter γ related to the extension ratio as $R_{lin}/R_{circ} = 2/\gamma$.*

| Buffer concentration (TBE) | Mean extension ratio | Mean STD ratio |
|---|---|---|
| *Experiment:* | | |
| 0.05X (100x100 nm$^2$) | 1.84 ± 0.07 | 2.64 ± 0.64 |
| 0.1X (100x150 nm$^2$) | 1.76 ± 0.06 | 2.57 ± 0.67 |
| 0.5X (100x150 nm$^2$) | 1.72 ± 0.05 | 2.45 ± 0.83 |
| 2.5X (100x150 nm$^2$) | 1.64 ± 0.07 | 2.26 ± 0.52 |
| Levy *et al.* (5X, 150x135 nm$^2$) | 1.54 ± 0.15 | |
| Levy *et al.* (5X, 215x155 nm$^2$) | 1.47 ± 0.15 | |
| *Theory:* | | |
| Odijk | 1.92-2.0 | 2-2.5 |
| Extended de Gennes | 1.59 | 2 |

**Table of Contents Graphic**

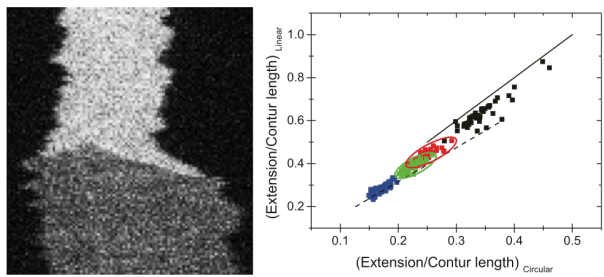

# Nanoconfined circular and linear DNA – equilibrium conformations and unfolding kinetics

Mohammadreza Alizadehheidari[a], Erik Werner[b], Charleston Noble[c,e], Michaela Reiter-Schad[c], Lena K. Nyberg[a], Joachim Fritzsche[d], Bernhard Mehlig[b], Jonas O. Tegenfeldt[e], Tobias Ambjörnsson[c], Fredrik Persson[f] and Fredrik Westerlund[a]*